\documentstyle[aps,pra,floats,epsf,multirow]{revtex}
\input texdraw

\def\la{\mathrel{\mathpalette\fun <}}
\def\ga{\mathrel{\mathpalette\fun >}}
\def\fun#1#2{\lower3.6pt\vbox{\baselineskip0pt\lineskip.9pt
  \ialign{$\mathsurround=0pt#1\hfil##\hfil$\crcr#2\crcr\sim\crcr}}}

\def\es{\nu_e\rightleftharpoons\nu_s}
\def\etau{\nu_e\rightleftharpoons\nu_\tau}

\def\mue{\nu_\mu\rightleftharpoons\nu_e}
\def\mus{\nu_\mu\rightleftharpoons\nu_s}
\def\muprimes{\nu_\mu^*\rightleftharpoons\nu_s}
\def\muprimee{\nu_\mu^*\rightleftharpoons\nu_e}

\def\mutau{\nu_\mu\rightleftharpoons\nu_\tau}
\def\nuebarnue{$\nu_e$/$\bar\nu_e$\ }
\def\me2{m_{\nu_e}^2}
\def\mmu2{m_{\nu_\mu}^2}
\def\mtau2{m_{\nu_\tau}^2}
\def\ms2{m_{\nu_s}^2}
\def\sinemu{\sin ^22\theta_{e\mu}}
\def\sines{\sin ^22\theta_{es}}

\def\sinmutau{\sin ^22\theta_{\mu\tau}}
\def\beq{\begin{equation}}
\def\eeq{\end{equation}}

\begin{document}

\draft
\title{The Increase in the Primordial $^4$He Yield
in the \\ Two-Doublet Four-Neutrino Mixing Scheme}

\author{Kevork~Abazajian, George M. Fuller and Xiangdong~Shi}
\address{Department of Physics, University of California,
San Diego, La Jolla, California 92093-0319}

\date{July 11, 2000}

\maketitle

\begin{abstract}
We assess the effects on Big Bang Nucleosynthesis (BBN) of lepton
number generation in the early universe resulting from the two-doublet
four-neutrino mass/mixing scheme. It has been argued that this
neutrino mass/mixing arrangement gives the most viable fit to the
existing data.  We study full $4 \times 4$ mixing matrices and show
how possible symmetries in these can affect the BBN $^4$He abundance
yields.  Though there is as yet no consensus on the reliability of BBN
calculations with neutrino flavor mixing, we show that, in the case
where the sign of the lepton number asymmetry is unpredictable, BBN
considerations may pick out specific relationships between mixing
angles.  In particular, reconciling the observed light element
abundances with a $\bar\nu_\mu \rightleftharpoons \bar\nu_e$ 
oscillation interpretation of LSND would allow unique new constraints
on the neutrino mixing angles in this model.
\end{abstract}
\bigskip


\section{Introduction}
In this paper we revisit the long standing issue of how neutrino
flavor mixing in the early universe might affect Big Bang
Nucleosynthesis (BBN). Though this is an old problem
\cite{early,Dolgov,Enqvist,Cline,Shi1}, with the advent of modern
experiments ({\it e.g.}, solar, atmospheric, and accelerator-based
oscillation experiments) we can hope to begin constructing the
neutrino mass/mixing matrix.  Here we adopt the leading model for
this, the two-doublet four-neutrino mass scheme \cite{Caldwell,bgg}.
We discuss this mixing in detail, going beyond recent good assessments
of this problem \cite{bellfootvolkas,okadayasuda,bilenkybbn}, to
examine what BBN considerations {\it ultimately} may be able to tell
us about parameters in this mixing scheme.

\begin{figure}
\centerline{\epsfxsize 6.5 in \epsfbox{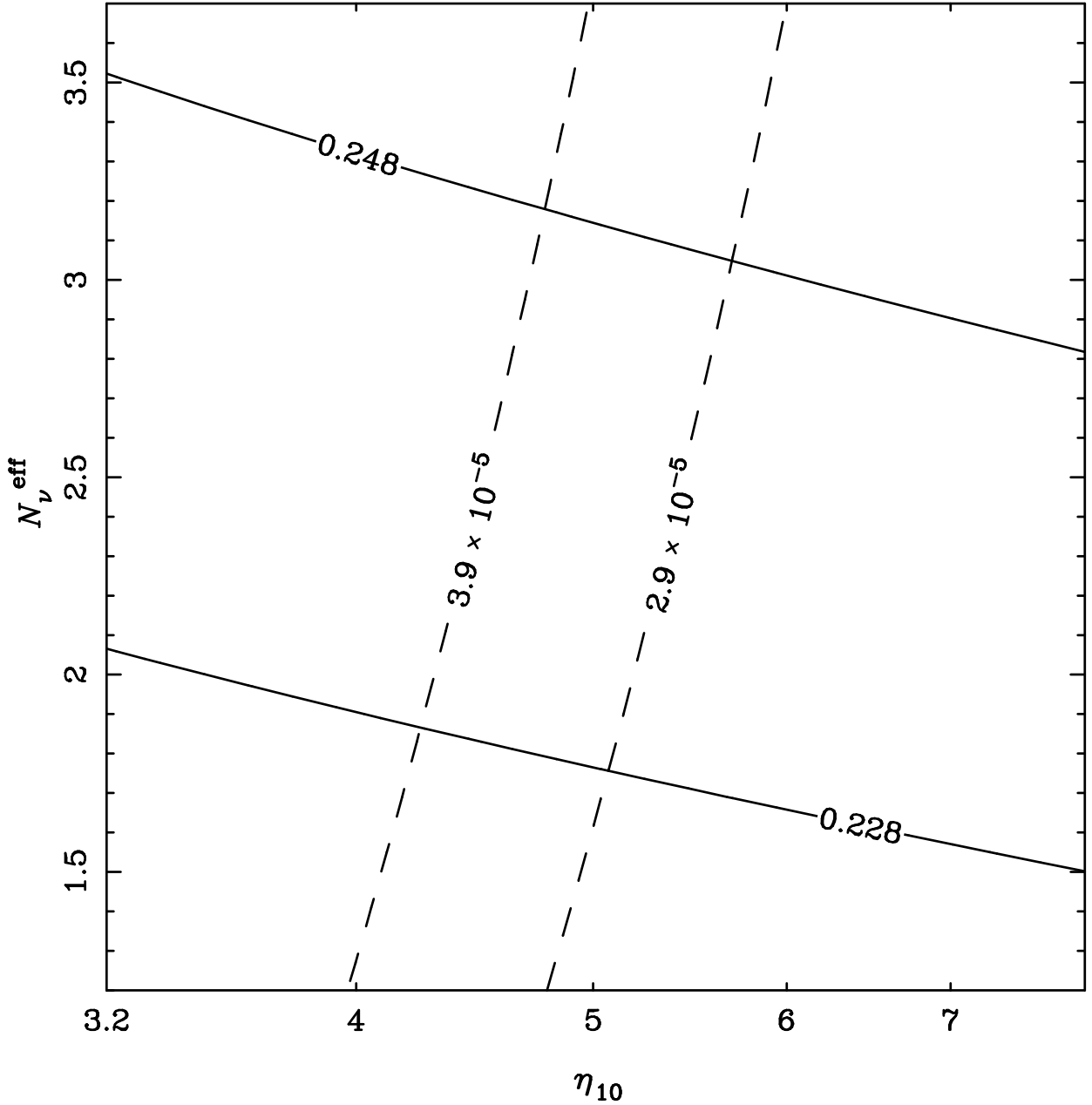}}
\vskip 1.5 cm
\caption[]{\small The limits in the effective number of light
neutrinos, $N_\nu^{\rm eff}$, in BBN for varying baryon-to-photon
ratio, $\eta_{10}\equiv \eta\ 10^{10}$.  The solid contours are 95\%
confidence limits on the inferred $^4$He mass fraction, $Y$
\cite{oss}, and the dashed contours are 95\% confidence limits on the
inferred relative abundance of $D/H$ \cite{burtyt}.
\label{nnu}
}
\end{figure}

Historical attempts \cite{early,Dolgov,Enqvist,Cline,Shi1} at
constraints on neutrino mixing are based on the argument that an
active-sterile neutrino mixing that is too large at (or prior to) the
BBN epoch will populate the sterile neutrino sea.  The resultant
increase in the total energy density of the universe at a given
temperature speeds up the Hubble expansion. In turn this leads to a
higher weak-freeze-out temperature and, consequently, it could lead to
a higher neutron-to-proton ratio at Nuclear Statistical Equilibrium
freeze-out \cite{Schramm} (however, the neutron-to-proton ratio is
determined not only by the expansion rate, see below).  Since
essentially all neutrons are incorporated into $^4$He in the early
universe, such a mixing yields a higher $^4$He abundance $Y$. This
potentially could contradict the observationally inferred primordial
$^4$He abundance. From the inferred abundances of $^4$He and $\rm
D/H$ there are strong constraints on the increase of the predicted
$Y$ from an increased energy density due to sterile neutrino
production.  These constraints are often translated into a limit
on the ``effective number of neutrinos'' of
\begin{equation}
1.7 \leq N_\nu^{\rm eff} \leq 3.2
\end{equation}
at 95\% confidence that is strictly limited by the baryon-to-photon
ratio, $\eta$, determined by the inferred relative abundance of
deuterium ($\rm D/H$) \cite{burtyt}.  This limit is shown in
Fig.~\ref{nnu}, where we have calculated $Y$ and $\rm D/H$ over a
range in $N_\nu^{\rm eff}$ and $\eta$. We have included corrections to
$Y$ derived from the zero and finite-temperature radiative, Coulomb
and finite-nucleon-mass corrections to the weak rates; order-$\alpha$
quantum-electrodynamic correction to the plasma density, electron
mass, and neutrino temperature; incomplete neutrino decoupling; and
numerical time-step effects \cite{lte}.  The observationally-inferred
primordial mass fraction we adopt is generous: $0.228 < Y < 0.248$
(95\% confidence range) \cite{oss}; and the inferred relative
abundance $\rm D/H$ we adopt is the well-established ${\rm D/H}
\approx (3.4 \pm 0.5) \times 10^{-5}$ \cite{burtyt}.  For arguments
and observational evidence on the reliability of the deuterium
abundance see Ref.~\cite{confbt}.  Similar limits on $N_\nu^{\rm eff}$
have been derived by other groups \cite{cardfull,Shi5,Mike}. These
limits preclude by a wide margin a fully populated sterile neutrino
sea ($N_\nu^{\rm eff} =4$) in the BBN epoch.  The concept of
$N_\nu^{\rm eff}$ (an expansion rate measure) as the sole determinant
of the $^4$He yield is a misleading and dangerous one.

A few $4\times 4$ models have been advanced as simultaneous
explanations of the atmospheric neutrino data, the solar neutrino
data, and the data from LSND \cite{Caldwell,bgg}. One hierarchical
scheme would have (near) maximal $\mutau$ mixing for the atmospheric
deficit, a $\nu_e\rightleftharpoons\nu_\mu$ mass splitting that gives
either a vacuum or MSW solar solution, and the LSND indication of a
large $\delta m^2$ is given by ``indirect'' mixing through a sterile.
This is Scheme I in Fig.~\ref{schemes}.  In this scheme, the mass
states most closely associated with the active neutrinos form a
triplet that has a significant mass splitting from the mostly sterile
mass state.  This scheme is ruled out by BBN because it requires large
$\mus$ and $\es$ mixing amplitudes to explain the LSND results through
the indirect conversion $\nu_\mu\rightarrow \nu_s \rightarrow \nu_e$
(and the $CP$ conjugate).  The required large mixing amplitudes for
$\mus$ and $\es$ populate the sterile sea in the early universe
through direct oscillation production of steriles
\cite{Dolgov,Enqvist,Cline,Shi1}.  This mass model is also disfavored
by a combined $4\times 4$ experimental data analysis \cite{bgg}.

An alternative two-doublet hierarchical scheme (Scheme II) has (near)
maximal $\mus$ mixing to fit the atmospheric data in Super-Kamiokande,
and $\etau$ mixing explaining the solar neutrino puzzle. This option
is excluded by BBN since the $\mus$ transformation is too large to be
compatible with $^4$He observations \cite{Shi1,Shi4}.  (Recently Foot
and Volkas \cite{Foot2} have argued that $\nu_\mu \rightleftharpoons
\nu_s$ maximal vacuum mixing for SuperK atmospheric neutrinos in a
related mass scheme {\it can} be reconciled with BBN limits.)
\footnote{There is still another loop-hole: if there is a pre-existing
lepton number asymmetry with a magnitude $\ga 10^{-5}$ during the BBN
epoch, the $\mus$ mixing can be suppressed \cite{FootVolkas}. This
assumption must then involve physics that is beyond neutrino mixing.}

Therefore, adopting the previously considered limits from BBN, we are
left only with the two-doublet four-neutrino mixing model, Scheme III.
The BBN effects of the model in Scheme III were considered in
Ref.~\cite{bellfootvolkas}.  In this paper we expand on the analysis
of the mixing and suggest how BBN could give potentially stricter
limits that those found in Ref.~\cite{bellfootvolkas}. (We note here
that the ``inverted scheme,'' where the $\nu_e/\nu_s$ doublet is more
massive than the $\nu_\mu/\nu_\tau$ doublet is not yet completely
ruled out by laboratory and astrophysical considerations.)

A new twist was added to the saga of neutrino-mixing in the early
universe when it was found that resonant active-sterile neutrino
transformation in the BBN epoch can alter neutrino energy spectra and
generate a lepton number asymmetry\cite{Foot1,Shi2}. This raises the
possibility that an asymmetry in \nuebarnue numbers could be
generated. In the mean time, their energy spectra may be modified so
that the weak reaction rates themselves may change, resulting in a
different neutron-to-proton ratio and a different $^4$He yield
$Y$\cite{Foot2,Shi3,ASF}.  It has been argued that a positive
\nuebarnue asymmetry behaves like a positive chemical potential for
$\nu_e$.  This would reduce $Y$, and by the same token a negative
asymmetry would increase it\cite{Foot2}.  However, this argument of a
direct asymmetry-$Y$ leverage relation is too naive in the context of
active-sterile neutrino mixing, and is in fact incorrect. This is
because the process of lepton number generation via resonant
active-sterile neutrino mixing potentially has a crucial and unique
feature\cite{Shi5,Shi2,Enqvist2,sorri,barifoot}: that the lepton
number asymmetry is first damped to essentially zero and then can
oscillate chaotically with an increasingly larger amplitude, until it
converges to a growing asymptotic value that is either positive or
negative.

However, the existence of ``chaoticity'' (unpredictability) in the
final sign of the lepton asymmetry, $L$, is controversial.  Indeed,
{\it all} claims ({\it e.g.} \cite{Foot1,Shi2}) for lepton number
generation in the early universe via neutrino mixing are now at odds
with at least one calculation (Dolgov, Hansen, Pastor and Semikoz
\cite{dh}).  Ref.~\cite{dh} uses a new formulation of the solution of
the neutrino energy density matrix evolution that finds a non-chaotic
generation of only a small and insignificant lepton asymmetry.  Refs.
\cite{Enqvist2} and \cite{sorri} recalculated the lepton number
generation found in Ref.~\cite{Shi2} and corroborated a random nature
to the sign of the $L$.  Ref.~\cite{sorri} showed that certain
over-simplifying approximations in Ref.~\cite{dh} may have
unphysically stabilized the evolution of the lepton number.
Ref.~\cite{barifoot} found a randomness in the sign of $L$, but only
for small non-phenomenological mixing angles, much smaller than those
that may be involved in neutrino oscillation solutions to the
experiments discussed here.  The oscillations in lepton number sign
seen in Ref.~\cite{barifoot} only occur below the precision of their
numerical solution.  One may call into question then the ability of
their numerical formulation to resolve oscillations occuring at larger
mixing angles which can be found in the more straight-forward
momentum-averaged solution of Refs.  \cite{Shi2,Enqvist2,sorri}.  The
momentum-averaged solution is valid at the relevant high-temperatures
where the instantaneous approximation to repopulation are good.  For
the LSND mass scale in Scheme III considered here,
$m_{\nu_\mu}^2-m_{\nu_e}^2 \approx m_{\nu_\mu}^2-m_{\nu_s}^2\sim 0.2$
to 10 eV$^2$ and a small effective mixing angle ($\sin^2\theta_{\rm
eff} > 10^{-10}$), the bulk of the active neutrinos undergo resonance
at a temperature $T\sim 10$ to 20 MeV where the instantaneous
approximation can be considered valid.

Given the controversy and disagreement among these different
calculations it is difficult to take any BBN-derived constraint with
confidence.  With this caveat in mind, we explore what constraints
{\it might} be possible if the lepton number generation magnitude is
as in Refs.~\cite{Foot1,Shi2}, but where chaoticity in lepton number
sign obtains for relevant parameters.

In a chaotic lepton number generation regime, the sign of the lepton
number asymmetry is independent of the initial conditions before the
amplification begins, and is exponentially sensitive to the parameters
involved during the chaotic oscillitory phase \cite{Shi2}.  In turn,
the sign of the asymmetry cannot correlate over a scale bigger than
the particle horizon \cite{Shi4}.

This causal structure of space-time can make it impossible to obtain a
universe with a uniform lepton number asymmetry. Instead, the lepton
number generating process gives rise to a universe with numerous
lepton number domains with similar lepton number magnitudes, albeit
different signs \cite{Shi4}. The size of each domain is less than the
horizon size at that epoch ($\sim 10^{10}$ cm at the weak-freeze-out
temperature, although the detailed geometric structure of the domains
at the BBN epoch depends on the manner in which domain ``percolation''
occurs). The distribution of domains with different signs is
completely random so that in total each sign occupies half of the
space.

The overall primordial $^4$He yield in such a universe must be an
average of $Y$ over domains with opposite \nuebarnue asymmetries.
Interestingly, this implies that the overall $Y$ is always {\it
larger} than that expected when there are no lepton asymmetries. This
is because the increase in $Y$ from a negative lepton asymmetry
generated by the resonant active-sterile neutrino mixing process is
always {\it bigger} than the decrease in $Y$ from a positive asymmetry
generated in the same process\cite{Shi3}.

In this paper we examine in detail how this causality effect operates
in the context of a {\it specific} four neutrino mixing scheme. There
are some surprises.

We can quantify these arguments for the two-doublet neutrino mixing
model (Scheme III).  In this model, the $\mue$ mixing that fits the
LSND data has a mass-squared-difference $\mmu2-\me2\sim 0.2$ to 10
eV$^2$ and an effective two-neutrino mixing angle satisfying
$\sinemu\sim 10^{-3}$ to $10^{-2}$\cite{LSND}; the $\es$ mixing that
solves the solar neutrino puzzle has either $\ms2-\me2\sim 10^{-5}$
eV$^2$ and an effective two-neutrino mixing angle satisfying
$\sines\sim 10^{-2}$ for the small mixing angle (SMA) MSW
(Mikheyev-Smirnov-Wolfenstein) solution or $\vert\ms2-\me2\vert\sim
10^{-10}$ eV$^2$ and $\sines\sim 1$ for the vacuum
solution \cite{Bahcall}; the Super-Kamiokande atmospheric neutrino
data are nicely fit by $\mutau$ mixing with
$\vert\mmu2-\mtau2\vert\sim 10^{-3}$ to $10^{-2}$ eV$^2$ and
$\sinmutau\sim 1$\cite{SuperK}.  The hierarchy in masses which emerges
from this mixing scheme includes an upper doublet of heavier
neutrinos, consisting of (almost) degenerate $\nu_\mu$ and $\nu_\tau$,
and a lower doublet of lighter neutrinos, consisting of slightly mixed
(in the case of the MSW solution) or (near) maximally mixed (in the
case of vacuum mixing) $\nu_e$ and $\nu_s$. The inter-doublet mixing
between $\nu_\mu$ and $\nu_e$ is small.  So also must be the
inter-doublet mixing between $\nu_\mu$ or $\nu_\tau$ and $\nu_s$, so
as to avoid conflicts with BBN \cite{Enqvist,Cline,Shi1}.

The large mixing angle (MSW) solution to the solar neutrino problem has
been previously disfavored by BBN \cite{Enqvist,Cline,Shi1}. We
consider the SMA MSW active-sterile neutrino mixing solar solution and
the vacuum active-sterile neutrino mixing solar solution. In Ref.\
\cite{Bahcall}, the vacuum sterile neutrino solar solution was
considered to be disfavored using the current combined solar neutrino
experiment rate data.  However, in Ref.\ \cite{goswami}, it was argued
that if only the solar Super-K spectrum or the seasonal solar data are
considered, then the vacuum active-sterile neutrino mixing solution
gives a better fit than a vacuum active-active neutrino mixing
solution.  Since the nature of neutrino mixing of solar neutrinos is
still not certain and since BBN considerations may prove to be
enlightening, we entertain the possibility of a vacuum
active-sterile neutrino mixing solution to the solar neutrino
deficit. The vacuum sterile neutrino mixing solution to the solar
neutrino problem was also considered in terms of a neutrino mass model
represented by the [SU(3)]$^3$ or $E^6$ groups in Ref.\
\cite{chmoh00}.

The interpretation of current experimental results usually is framed
in terms of an effective two neutrino mixing scenario (i.e., in terms
mass-squared-differences and mixing angles) for each experimental
situation.  This model is approximately valid in the two-doublet
four-neutrino mixing scheme because, as a result of the mass
hierarchy, one two-species mixing dominates in each of the above
experiments. It is, however, more informative to employ the full
$4\times 4$ mixing matrix in our discussion. In the next section we
will briefly review what has been learned about this mixing matrix
from the current experiments. We will then proceed to consider BBN
$^4$He synthesis in the presence of hierarchical four-neutrino mixing
schemes. From the $^4$He yield we infer potential new constraints on
the inter-doublet mixing matrix elements between active neutrinos and
the sterile neutrinos. In section III, we will summarize our results.

\section{The Hierarchical Four-Neutrino Scheme
and the Primordial $^4$He Abundance}

We adopt the convention of employing Greek indices to denote flavor
eigenstates $\nu_s$, $\nu_e$, $\nu_\mu$ and $\nu_\tau$, and employing
Latin indices to denote mass eigenstates $\nu_0$, $\nu_1$, $\nu_2$ and
$\nu_3$. The two bases are related by a unitary transformation $U$:
\beq 
\nu_\alpha=\sum_{i=0,3}U_{\alpha i}\nu_i.  
\eeq 
The mass matrix in the flavor basis is then 
\beq
M_{\alpha\beta}=\sum_{k=0,3}\sum_{l=0,3} U^*_{\alpha
k}m_k\delta_{kl}U^\dagger _{l\beta}, 
\eeq 
where $m_k$ are the mass eigenvalues, and $\delta_{kl}$ are the
Kronecker deltas.  In the scheme considered here, $m_0,\,m_1\ll
m_2,\,m_3$.

The full expression for $U$ can be found, for example,
in Eq.~(8) of Ref.~\cite{Barger1}. It has 12 degrees 
of freedom, parametrized by 6 rotation angles $\theta_{ab}$,
and 6 $CP$-violating phases $\phi_{ab}$. The parameters
are symmetric in indices $a$ and $b$, which run from 0
to 3 ($a<b$). We follow convention and write $s_{ab}\equiv
\sin\theta_{ab}e^{i\phi_{ab}}$ and $c_{ab}\equiv\cos\theta_{ab}$.
Because the inter-doublet mixing is small,
$\vert U_{s2}\vert^2$, $\vert U_{s3}\vert^2$,
$\vert U_{e2}\vert^2$, $\vert U_{e3}\vert^2$, 
$\vert U_{\mu 0}\vert^2$, $\vert U_{\mu 1}\vert^2$,
$\vert U_{\tau 0}\vert^2$ and $\vert U_{\tau 1}\vert^2$
are small, implying that
$\vert s_{02}\vert,\,\vert s_{03}\vert,\,
\vert s_{12}\vert,\,\vert s_{13}\vert$ are small.
In fact, the Bugey result can be translated into a limit
$\vert s_{12}\vert,\,\vert s_{13}\vert\la 0.1$\cite{Barger1}.
The assumption that $\nu_\mu$ and $\nu_\tau$ are (nearly)
maximally mixed yields $c_{23}\sim \vert s_{23}\vert\sim 1/\sqrt{2}$.
Furthermore, the LSND result suggests that it is likely that
$\vert s_{12}\vert,\,\vert s_{13}\vert\sim 0.1$\cite{Barger1}.
Finally and obviously, we should have
$c_{02}\sim c_{03}\sim c_{12}\sim c_{13}\sim 1$.

To leading order in $s_{02}$, $s_{03}$, $s_{12}$,
and $s_{13}$, we have \cite{Barger1}
\beq
U \approx \left( \begin{array}{cccc}
c_{01}  & s_{01}^* & \ s_{02}^* \ & \ s_{03}^* \ \\
&&&\\
-s_{01} & c_{01}   & \ s_{12}^* \ & \ s_{13}^* \ \\
&&&\\
-c_{01}(s_{23}^*s_{03}+c_{23}s_{02}) & -s_{01}^*(s_{23}^*s_{03}+c_{23}s_{02})
                                     & \ c_{23} \ & \ s_{23}^* \ \\
+s_{01}(s_{23}^*s_{13}+c_{23}s_{12}) & -c_{01}(s_{23}^*s_{13}+c_{23}s_{12})
                                     &            &              \\
&&&\\
c_{01}(s_{23}s_{02}-c_{23}s_{03})    & s_{01}^*(s_{23}s_{02}-c_{23}s_{03})
                                     & \ -s_{23}\ & \ c_{23}   \ \\
-s_{01}(s_{23}s_{12}-c_{23}s_{13})   & +c_{01}(s_{23}s_{12}-c_{23}s_{13})
                                     &            &              \\
&&&\\
\end{array} \right) \,.
\label{approxU}
\eeq
In Ref. \cite{Barger1} this matrix is discussed and is shown to be
unitary to second order in the LSND mixing angle.

In a simplifying approximation, one can take all mixing angles except
$\theta_{01}$, $\theta_{12}$ and $\theta_{23}$ to be zero, 
ignore the $CP$ violating phases, and take the Super-K
associated mixing to be maximal, $c_{23}\simeq 1/\sqrt{2}$.  In
this case, the transformation matrix becomes
\begin{equation}
U \approx \pmatrix{
c_{01}  & s_{01} & 0 & 0 \cr 
-s_{01} & c_{01} & s_{12} & 0 \cr 
s_{01}s_{12}/\sqrt{2} & -c_{01}s_{12}/\sqrt{2} & 1/\sqrt{2} 
&  1/\sqrt{2} \cr 
-s_{01}s_{12}/\sqrt{2} & c_{01}s_{12}/\sqrt{2} & -1/\sqrt{2} 
& \ 1/\sqrt{2} \cr
} \, ,
\label{cfqU}
\end{equation}
which is (approximately) unitary for nearly maximal mixing,
$c_{23}\simeq 1/\sqrt{2}$.

We can define a linear row transformation \cite{cfq,bf}
\begin{equation} 
  | \nu_\mu^* \rangle \equiv {|\nu_\mu\rangle - |\nu_\tau
    \rangle\over{ \sqrt{2}}},
\end{equation} 
and 
\begin{equation} 
  | \nu_\tau^* \rangle \equiv {|\nu_\mu\rangle + |\nu_\tau \rangle
    \over{ \sqrt{2}}},
\end{equation}
such that Eq.\ (\ref{cfqU}) becomes 
\begin{equation} U' \approx \pmatrix{ 
c_{01} & s_{01} & 0 & 0 \cr 
-s_{01} & c_{01} & s_{12} & 0 \cr 
s_{01} s_{12} & -c_{01} s_{12} & 1 & 0 \cr
0 & 0 & 0 & 1 \cr 
} \, ,
\label{cfqU2} 
\end{equation}
and 
\begin{equation}
\pmatrix{
|\nu_s \rangle \cr
|\nu_e \rangle \cr
|\nu_\mu^* \rangle \cr
|\nu_\tau^* \rangle \cr
}
= U'
\pmatrix{
|\nu_0 \rangle \cr
|\nu_1 \rangle \cr
|\nu_2 \rangle \cr
|\nu_3 \rangle \cr
}\,.
\end{equation}
Here, we can see that the fourth state $|\nu_\tau^*\rangle$ is a mass
eigenstate.  If there are no lepton asymmetries generated by a
mechanism other than neutrino mixing, then the muon- and tau-neutrino
flavors see the same matter effects (that is the same thermal and
fermion potentials) throughout their evolution.  The state
$|\nu_\mu^*\rangle$ is mixed with the sterile and electron neutrino,
and will undergo resonant MSW transformation under the appropriate
conditions.  However, $|\nu_\tau^*\rangle$ will pass through
resonances unchanged.  This reduces the $4\times 4$ mixing to
essentially a $3\times 3$ evolution, at least as far as MSW resonances
are concerned.\footnote{Note that Ref.~\cite{bellfootvolkas} also
points out that one ``linear combination'' of $\nu_\mu$ and $\nu_\tau$
``oscillates'' with $\nu_s$ while the other decouples.}  It should be
noted that the mixing matrix discussed here uses the rotation order
convention $U=R_{23}R_{13}R_{03}R_{12}R_{02}R_{01}$, while Caldwell,
Fuller and Qian use $U_{\rm CFQ} = R_{23}R_{01}R_{12}$ \cite{cfq}.
The different unitary mixing matrices are physically equivalent under
the same approximations, and exhibit the same decoupling seen in
Eq.\ (\ref{cfqU2}).

The matter effects present at the epoch of BBN comprise two pieces: a
part due to a finite-temperature thermal bath, and a part due to a
possible lepton number asymmetry in the active neutrino sectors (the
baryon number asymmetry and the associated electron-positron asymmetry
is too small to play a significant role in the neutrino mixing). For
two-neutrino mixing, the effective matter mass-squared-difference is
\beq {\delta m^2_{\alpha\beta}}^{\rm (eff.)}=\left\{
\left(m_{\nu_\alpha}^2-m_{\nu_\beta}^2\right)^2
\sin^22\theta_{\alpha\beta}
+\left[\left(m_{\nu_\alpha}^2-m_{\nu_\beta}^2\right)
\cos2\theta_{\alpha\beta}+2EV_{\alpha\beta}^T
+2EV_{\alpha\beta}^L\right]^2\right\}^{1/2} \eeq and the effective
mixing angle satisfies \beq \tan\theta_{\alpha\beta}^{\rm (eff.)}
={(m_{\nu_\alpha}^2-m_{\nu_\beta}^2)\sin 2\theta_{\alpha\beta}\over
(m_{\nu_\alpha}^2-m_{\nu_\beta}^2)\cos2\theta_{\alpha\beta}+2EV_{\alpha\beta}^T
+2EV_{\alpha\beta}^L}.  \eeq
Here we use $m_{\nu_\alpha}$ to denote the mass eigenstate most
closely associated with a neutrino of flavor $\alpha$; $E$ is the
neutrino energy; $V_{\alpha\beta}^T$ is the effective potential due to
the finite-temperature part of the matter effect, and
$V_{\alpha\beta}^L$ is the effective potential due to the lepton
number asymmetry.  Both potentials vary with temperature $T$.

We will first consider the case that the lepton number asymmetry
is negligible, $2EV_{\alpha\beta}^L\ll 2EV_{\alpha\beta}^T,\,
(m_{\nu_\alpha}^2-m_{\nu_\beta}^2)\cos2\theta_{\alpha\beta}$.
Mixings between active neutrino species alone do not modify the
$^4$He synthesis because active neutrinos share the same number
density distribution in the BBN epoch. (This is not rigorously
true because electron/positron annihilation overpopulates
\nuebarnue slightly but its impact on the $^4$He yield is less
than 0.1\% \cite{Dodelson}.)  Mixings between active neutrinos
and sterile neutrinos, however, convert active neutrinos and so
populate initially unoccupied sterile neutrino states. These 
mixings can therefore affect the energy spectra of active 
neutrinos, as well as the $^4$He yield.  In particular, MSW
resonances, at which the local effective mixing reaches maximal
values, can occur between an active
neutrino species $\nu_\alpha$ and the sterile neutrino species
when $(m_{\nu_\alpha}^2-m_{\nu_s}^2)\cos2\theta_{\alpha s}
+2EV_{\alpha s}^T=0$.  Since
\beq
V_{\alpha s}^T \approx -A{n_{\nu_\alpha}+n_{\bar\nu_\alpha}\over
n_\gamma}G_F^2ET^4,
\label{VT}
\eeq
where $A\approx 105 (30)$ for $\alpha=e\,(\mu,\tau)$, this
resonance condition is met when the temperature of the universe is 
\beq
T_{\rm res}\approx T_0\,\left({E\over T}\right)^{-1/3}
\left\vert{(m_{\nu_\alpha}^2-m_{\nu_s}^2)\cos2\theta_{\alpha s}
\over 1{\rm eV}^2}\right\vert^{1/6},
\label{Tres}
\eeq
where $T_0\approx 19(22)$ MeV for $\alpha=e\,(\mu,\tau)$.

Therefore, for a two-family mixing between $\nu_\mu^*$ and $\nu_s$,
with $m_{\nu_\mu^*}^2-m_{\nu_s}^2 \approx
m_{\nu_\tau^*}^2-m_{\nu_s}^2\sim 0.2$ to 10 eV$^2$ and a small mixing
angle, the bulk of the active neutrinos undergo resonance at a
temperature $T\sim 10$ to 20 MeV.  This is long before $\nu_\mu^*$ and
$\nu_\tau^*$ decouple thermally/chemically from the thermal
background.  As discussed briefly earlier, the
$\nu_\mu^*\rightleftharpoons\nu_s$ mixing potentials behave
identically to the $\nu_\tau^*\rightleftharpoons\nu_s$ in the BBN
epoch, and they undergo the same scattering and collision evolution.
Thus, the approximations giving Eq.\ (\ref{cfqU2}) effectively
``decouple'' $\nu_\tau^*$, with only $\nu_\mu^*$ going through
resonances.  In the following discussion, $\nu_\tau^*$ decoupling can
be assumed; however, even if the approximations leading to Eq.\
(\ref{cfqU2}) are invalid so that angles other than $\theta_{01},
\theta_{12}$ and $\theta_{23}$ are non-zero, the following discussion
is still relevant for the standard $\nu_\mu$ and can be extended to
$\nu_\tau$.

Resonances may also occur for $\es$ mixing with
$m_{\nu_e}^2-m_{\nu_s}^2\sim 10^{-10}$ eV$^2$ (a possible vacuum
solution for the solar neutrino problem). But the resonance
temperature is $\la 0.01$ MeV, which corresponds to an epoch long
after the weak-decoupling of neutrinos. Resonant active-sterile
neutrino mixing within the lower doublet in this case then cannot
influence BBN.  Therefore, the BBN constraints on mixing models
involving a vacuum solar solution are less stringent.

In the two-doublet neutrino mixing schemes considered here, the
$\muprimes$ channel is essentially decoupled (at or near its resonance
temperature) from the other mixings within the four species
family. Therefore, we can take the two-neutrino mixing picture as
applicable. This is so because while the $\muprimes$ channel is
matter-enhanced at its resonance, the other mixings are not, or even
are suppressed by the matter effects by a factor of $\vert
(m_{\nu_\alpha}^2-m_{\nu_\beta}^2)/2EV_{\alpha\beta}^T\vert^2$ with
respect to their vacuum mixing amplitude.  For example, the
inner-lower-doublet $\es$ mixing is suppressed by a factor $\sim 10^9$
for the SMA MSW mixing solar neutrino solution or $\sim 10^{19}$ for
the vacuum mixing solar neutrino solution, because $2EV_{es}\sim
8EV_{\mu s}\approx 4(\mmu2-\ms2)$.  Also, since $2EV_{\mu e}=2E(V_{\mu
s}-V_{e s})\approx -6EV_{\mu s} \approx 3(\mmu2-\ms2)\approx
3(\mmu2-\me2)$, the inter-doublet $\muprimee$ mixing is suppressed by
a factor $\sim 10$ with respect to its already small vaccuum mixing
amplitude ($\la 10^{-2}$).  Therefore, it is safe to employ the
two-family mixing picture to investigate the $\muprimes$ channel at or
near its resonance.

As pointed out in several previous papers, there are two possible
consequences of a resonant active-sterile neutrino mixing: (1) the
total neutrino energy density at a given temperature increases if
neutrino pair production is still effective in replenishing the
converted active neutrinos; (2) a lepton number asymmetry may be
generated in the active neutrino sector from initial small and
negligible statistical fluctuations during the resonant
active-to-sterile neutrino conversion process.  In either the
nonresonant or resonant mixing case, the limit on the total neutrino
energy density from the primordial $^4$He abundance puts the following
constraint on the parameters of active-sterile neutrino mixing
\cite{Enqvist}:
\begin{equation}
(m_{\nu_\alpha}^2-m_{\nu_s}^2)\sin^42\theta_{\alpha s}\la
\left\{\begin{array}{ll} 10^{-9}\,{\rm eV^2} & \mbox{if
$\nu_\alpha=\nu_e$;}\\ 10^{-7}\,{\rm eV^2} & \mbox{if
$\nu_\alpha=\nu_\mu,\,\nu_\tau$.}
\end{array}\right.
\end{equation}

In the resonant case, the increase in the neutrino energy density is
significant to BBN when the resonance is adiabatic and occurs before
chemical decoupling of the active neutrino (about 5 MeV for
$\nu_\mu^*$).  The lepton number asymmetry generated from an
initially very small asymmetry by the resonant active-sterile neutrino
mixing process is significant to BBN if part of the asymmetry resides
in the $\nu_e/\bar\nu_e$ sector and is of order $\ga 0.01$.  In the
two-doublet neutrino model, this is achieved by having a resonant
$\muprimes$ mixing generate $L_{\nu_\mu^*}$, and having a
resonant $\muprimee$ mixing transfer part of $L_{\nu_\mu^*}$ into
$L_{\nu_e}$.  

In Ref.~\cite{Shi3}, we calculated the change in the primordial $^4$He
abundance due to such a process.  In the regime where $10^{-1}$
eV$^2\la m^2_{\nu_\mu^*}-m^2_{\nu_s}\la 10$ eV$^2$ and
$\sin^22\theta_{\mu^* s}\ga 10^{-10}$, the mixing angles are
large enough to generate a lepton number asymmetry\cite{Shi2}. Mixings
with smaller mixing angles cannot have a material effect on BBN.
In regions of the universe where $L_{\nu_\mu^*}$ and in turn
$L_{\nu_e}$ is positive, the n$\rightarrow$p rate is enhanced while
the p$\rightarrow$n rate is reduced.  This change in
n$\rightleftharpoons$p rates tends to lower the neutron-to-proton ratio
and consequently $Y$.

On the other hand, the increased neutrino energy density from the
active-sterile neutrino mixing before neutrino chemical decoupling
always tends to increase $Y$.  The overall result is a decrease in
$Y$, as the former effect ($\nu_e/\bar\nu_e$ asymmetry) dominates. In
places where $L_{\nu_\mu^*}$ and $L_{\nu_e}$ are negative, however,
the n$\rightleftharpoons$p rates are changed so as to increase $Y$.  This
is, of course, in addition to the increase in $Y$ from the energy
density effect.

\begin{figure}
\center{\epsfxsize 5.5truein \epsfbox{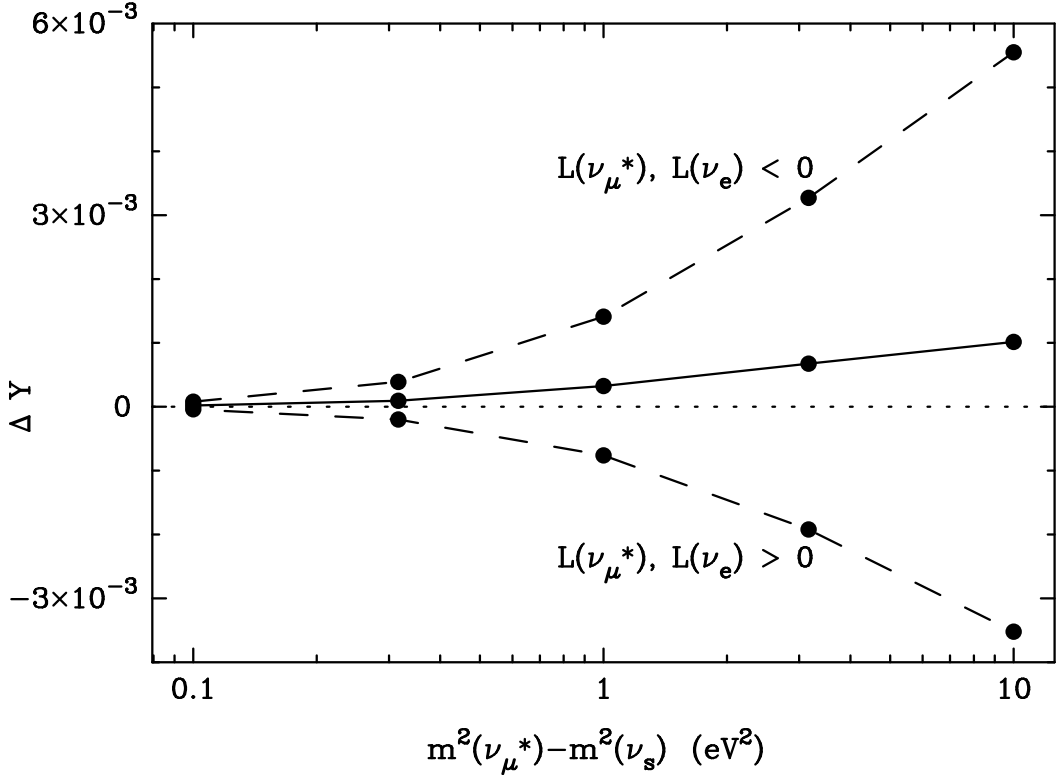}}
\caption[]{\small 
The increase in the primordial $^4$He yield in the
two-doublet mass scheme, as a function of the inter-doublet
mass-squared-difference. The mixing amplitude between $\nu_s$ and the
$\nu_\mu$-$\nu_\tau$ doublet is assumed to be not too small, $\ga
10^{-10}$.  The dashed curve: the increase in $Y$ in individual
domains. The solid curve: the increase in $Y$ averaged over positive
and negative lepton number domains.
\label{dy}
}
\end{figure}

When averaged over the positive domains and the negative domains, the
net $Y$ turns out to be consistently larger than that predicted by the
standard BBN picture assuming no neutrino mixing. This is because the
increase in $Y$ in negative domains is never compensated by the
decrease in $Y$ in positive domains, as shown in Fig.~\ref{dy}
\cite{Shi2,Shi4}.  

This result is rather different from that in Ref.
\cite{bellfootvolkas} and given the controversy in the calculations
alluded to above, it is difficult to say which, if either, is correct.
Part of the discrepancy may be because the effect of energy density
increases on $Y$ were not appropriately taken into account in
Ref.~\cite{bellfootvolkas}.  A simple average of the change in $Y$ in
our calculation is not beyond observational uncertainty bounds. The
average $\Delta Y$ is always less than $\sim 0.001$, or, equivilantly,
$N_\nu^{\rm eff}$ is always less than 3.08.  If a definitive,
confident solution to lepton number generation by neutrino mixing in
the early universe were to show unambiguously that the sign of the
neutrino asymmetry for this specific range of neutrino mixing
parameters is positive (negative) then the predicted change to $Y$
would follow the lower (upper) curve. The positive lepton number
result alone does not exceed the observational bounds of
Fig.~\ref{nnu}.  The observations more greatly constrain
increases to $Y$.  A negative lepton number result alone therefore
creates a $\Delta Y$ that is too large for $\delta m_{\mu^*e}^2 \ga
2.5 \rm\, eV^2$.

The averaged $Y$ is only a lower limit to the actual $Y$ in the
two-doublet neutrino model with chaotic lepton number generation.
This is because, as first pointed out by Shi and Fuller \cite{Shi4},
additional increases in $Y$ may arise from an extra channel for
increasing the total neutrino energy density. In the case where the
sign of the lepton number generated is undetermined, the extra channel
for sterile neutrino production results from the lepton number
gradients between domains of regions with opposite lepton number sign
(which are sub-horizon scale at the BBN epoch). The domain boundaries
can meet the conditions for resonant conversion of not only
$\nu_\mu^*$ to $\nu_s$ (and $\bar\nu_\mu^*$ to $\bar\nu_s$) but also
$\nu_e$ to $\nu_s$ (and $\bar\nu_e$ to $\bar\nu_s$).  Therefore, the
sterile neutrino sea is not only populated within domains by the
resonant neutrino mixing that drives the lepton number generation in
the first place, but can also be populated at domain boundaries by the
same resonant neutrino mixing as well as other active-sterile neutrino
mixings.

To avoid fully thermalizing the sterile neutrinos, therefore,
requires that this extra channel of sterile neutrino
production be suppressed. In other words, all resonances of active-sterile
neutrino conversion have to be non-adiabatic at domain boundaries.
This yields another limit for the two-doublet neutrino scheme\cite{Shi4}:
\beq
\begin{array}{ll}
\sin^22\theta_{\mu^* s}\la 10^{-10}, & \mbox{\small\ \ for an MSW solution to
                                        the solar neutrino problem;}\\
\left(m_{\nu_\mu^*}^2-m_{\nu_s}^2\right)
\sin^22\theta_{\mu^* s}\la 10^{-4}\,{\rm eV}^2, & \mbox{\small\ \ for a vacuum 
                                             solar neutrino solution.}
\end{array}
\label{stringentlimit}
\eeq 
Beyond these limits, $Y$ increases by at least 0.013 due to a
fully populated sterile neutrino sea resulting from active-sterile
neutrino conversions at domain boundaries.

These limits are summarized in Table 1.
Because of the decoupling of various mixings, the above
limits can be directly translated into constraints
\beq
\begin{array}{ll}
\vert s_{02}\vert,\,\vert s_{03}\vert\la 10^{-5}, 
                             & \mbox{\small\ for a SMA MSW solar
                                             neutrino solution}\\
        &\mbox{\small\  or when $m_{\nu_\mu^*}^2-m_{\nu_e}^2\ga 4$ eV$^2$;}\\
\vert s_{02}\vert,\,\vert s_{03}\vert\la 10^{-2}
\left[\left(m_{\nu_\mu^*}^2-m_{\nu_s}^2\right)/1\,{\rm eV}^2\right]^{-1/2},
                                           & \mbox{\small\ for a vacuum 
                                             solar neutrino solution}\\
        &\mbox{\small\ and $m_{\nu_\mu^*}^2-m_{\nu_e}^2\la 4$ eV$^2$.}
\end{array}
\label{stringentlimit2}
\eeq 
The constraint on the large mass difference is the result of the
fact that for $m_{\nu_\mu^*}^2-m_{\nu_e}^2\ga 4$ eV$^2$, domains of
$L_{\nu_\mu^*}$ will facilitate $\nu_\mu^*\rightleftharpoons \nu_s$
population of the sterile sea across domain boundaries.  These
constraints imply that the inter-doublet mixing elements of $\nu_\mu$
or $\nu_\tau$ with $\nu_s$ ($s_{02}, s_{03}$) are $\sim 10^4$ times
smaller than the elements associated with mixing with $\nu_e$
($s_{12}, s_{13}$), if the solar neutrino problem has its roots in
$\es$ mixing with $m_{\nu_s}^2-m^2_{\nu_e}\sim 10^{-5}$ eV$^2$ (the
SMA MSW solution). However, the mixings between the upper doublet
neutrinos and the lower doublet $\nu_e$ and $\nu_s$ are within an
order of magnitude if a vacuum ``just so''
$\nu_e\rightleftharpoons\nu_s$ mixing with $\vert
m_{\nu_s}^2-m^2_{\nu_e}\vert \sim 10^{-10}$ eV$^2$ explains the solar
neutrino data.

\begin{figure}
\center{\epsfxsize 6.2truein \epsfbox{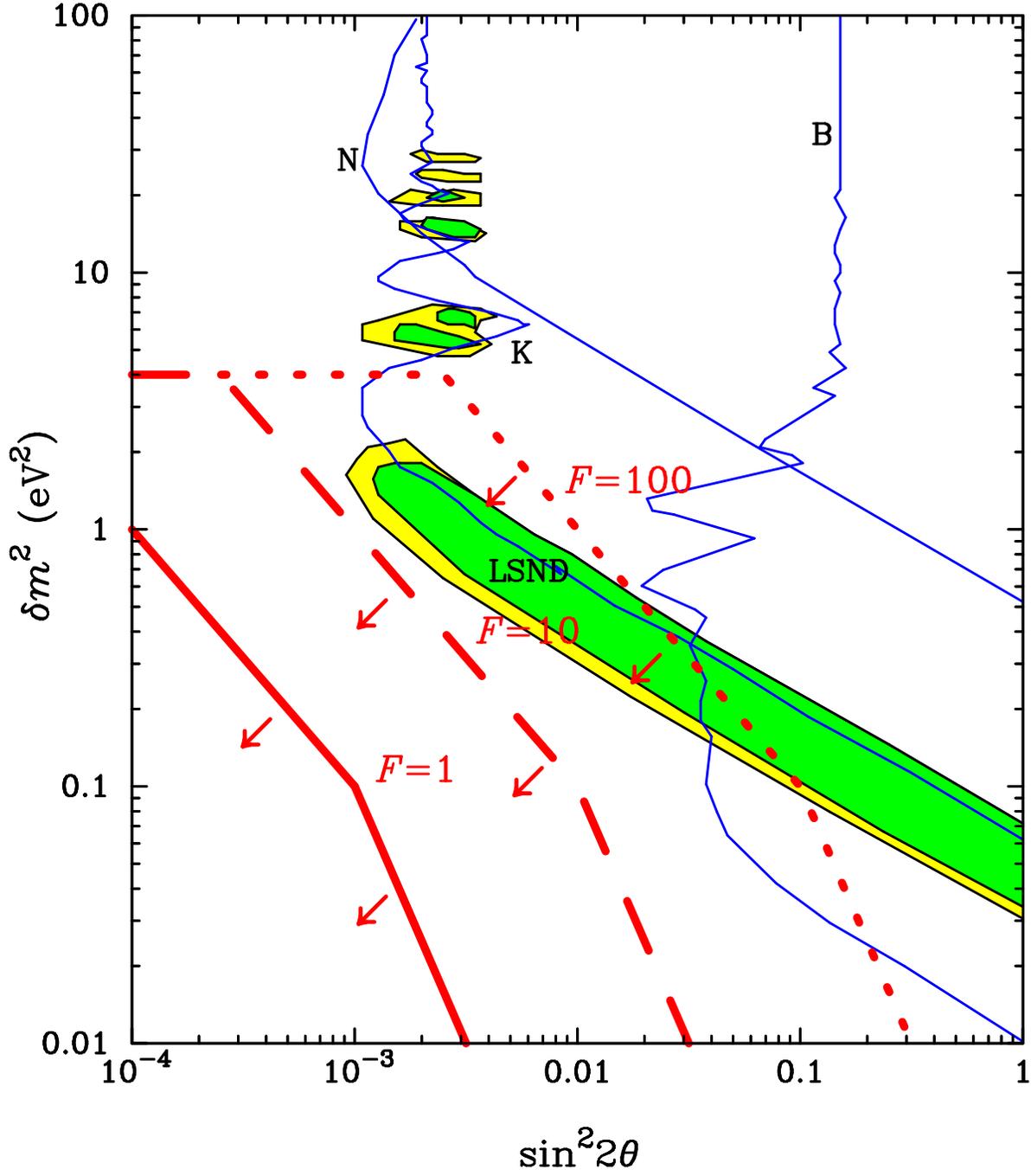}}
\vskip 1cm
\caption[]{\small The effective $\nu_\mu\rightleftharpoons\nu_e$
mixing parameters suggested by BBN and LSND for varying values of the
asymmetry factor, $F$. Shown are the 90\% and 95\% {\it CL} limits for
LSND.  Curves labelled K, B, and N are the 90\% {\it CL} limits from
KARMEN2, Bugey, and NOMAD, respectively. Experimental confidence
regions are adapted from Ref.\ \cite{eitel}.
\label{lsnd}
}
\end{figure}

Let us now confine our attention to the latter scheme, where $m^2_{\nu_\mu^*}
-m^2_{\nu_e}$ is $\la 4$ eV$^2$ and the
solar neutrino deficit is explained by (nearly) maximally mixed
$\es$ vacuum \lq\lq just so\rq\rq\ oscillation with $\delta m^2_{es}
\sim 10^{-10}$ eV$^2$. This scheme can produce a significant
increase in $Y$ when the values of the inter-doublet mass splitting
are in the range that is required to explain the LSND results. 
The primordial $^4$He yield in this scenario is sensitive to the
relative level of mixing for $\nu_{\mu,\tau}\rightleftharpoons\nu_e$ and
for $\nu_{\mu,\,\tau}\rightleftharpoons\nu_s$. For example, we can
define a factor
\begin{eqnarray}
F&=&{\sin^22\theta_{\mu e}\over\sin^22\theta_{\mu s}}\cr
&=& {\left|s_{12} c_{23} + s_{13} s_{23}^*\right|^2 \over
{\left|s_{02} c_{23} + s_{03} s_{23}^*\right|^2}},
\end{eqnarray}
where $\theta_{\mu e}$ and $\theta_{\mu s}$ are the effective
two-neutrino vacuum mixing angles (to be constructed from the matrix
elements in Eq.\ (\ref{approxU})) corresponding to
$\nu_\mu\rightleftharpoons\nu_e$ and $\nu_\mu\rightleftharpoons\nu_s$,
respectively. Figure \ref{lsnd} then shows limits from the observed
$^4$He abundance for three cases: (1) $F=1$; (2) $F=10$; and (3)
$F=100$. More specifically, if the atmospheric neutrino problem
solution is maximal $\mutau$ mixing (and non-$CP$ violating), then $F$
is just the ratio of inter-doublet mixing angles
\begin{equation}
F \simeq {\left|s_{12} + s_{13} \right|^2 \over
{\left|s_{02} + s_{03}\right|^2}}.
\end{equation}
It is noteable that a novel solution to r-process nucleosynthesis in
Type II supernovae by Caldwell, Fuller and Qian \cite{cfq} involves a
$4\times 4$ model where only $s_{12}$ is non-vanishing in the above
expression for $F$.  In this case, effectively $F\rightarrow\infty$,
indicating an exceptionally high degree of ``symmetry.''  (Here by
``symmetry'' we mean that only one mixing angle governs the
inter-doublet mixing, not four; however, this is very asymmetric as far
as $F$ is concerned!) Both this solution to r-process nucleosynthesis
and the domain-conversion-based BBN considerations discussed in this
paper favor a large $F$.  The mixing matrix in Ref.~\cite{cfq}, which
is related to that shown in Eq.\ \ref{cfqU2}, exhibits the symmetry
that allows a decoupling of a neutrino state ($|\nu_\tau^*\rangle$)
from MSW resonant evolution.  For mass Scheme III to both account for
the LSND signal (whose two-neutrino parameter space is shown in
Fig.~\ref{lsnd} \cite{eitel}) and be consistent with
domain-conversion-driven BBN effects, the difference in magnitude of
the inter-doublet mixing angles must be large (a large value of $F$).

If one were to entertain models where $F$ was not large ({\it i.e.},
comparable values for $\theta_{12}, \theta_{13}, \theta_{02}$ and
$\theta_{03}$), then already we can see from Fig.~\ref{lsnd} that the
LSND data compatible with smaller $F$ all lies near
$m^2_\mu-m^2_e\approx 1$ eV$^2$ and $\sin^22\theta_{\mu e}\approx
10^{-3}$.  Concommitantly, future experiments such as BooNE
\cite{boone} might indicate mixing parameters which fall outside this
limit.  Assuming our BBN domain-conversion-driven considerations are
correct, this would argue strongly for large $F$.

This would be a remarkable outcome.  Naively, one might assume that,
{\it e.g.}, the BooNE experiment measures only the effective
two-neutrino mixing $\theta_{\mu e}$.  However, when combined with
other experiments and the BBN physics adopted here, it is evident that
many other mixing matrix elements (those in $F$) are also probed.

\section{Summary}
We have explored the possible change in the primordial $^4$He
abundance in the currently favored two-doublet four-neutrino
($\nu_\tau$-$\nu_\mu$/$\nu_e$-$\nu_s$) mixing scheme proposed to
simultaneously explain the current neutrino experiments.  Though
definitive calculations of matter-enhanced neutrino conversion effects
are elusive at present (in the eyes of some), here we have adopted the
set of calculations which could give the tightest constraints on the
neutrino mass and mixing matrix.  We do this in the spirit of
determining what may be possible.  When we analyze the BBN effects in
the context of full four-neutrino mixing we find some remarkable
hints. Namely, we find that putative limits on the matrix elements
governing $\nu_{\mu,\tau} \rightleftharpoons \nu_e$ and
$\nu_{\mu,\tau} \rightleftharpoons \nu_s$ could be very restrictive.

We have found that these limits strongly depend on the mixing between
$\nu_e$ and $\nu_s$: the limits are exceptionally strong if the $\es$
mixing parameters are in the range of the SMA MSW solution of the
solar neutrino problem; but they are much less so, allowing the
$\nu_\mu,\,\nu_\tau\rightleftharpoons\nu_s$ mixing to be at the same
level as the $\nu_\mu,\,\nu_\tau\rightleftharpoons\nu_e$ mixing, if
the $\es$ mixing parameters lie at the vacuum \lq\lq just so\rq\rq\
solution region the solar neutrino problem. (Note: Since the
submission of this paper there has been recent evidence from
Super-Kamiokande that disfavors all sterile neutrino solutions to the
solar neutrino problem \cite{suzukinu2000}.  This emphasizes BBN
constraints on any four-neutrino models.) In addition, we have found
that if $m^2_{\nu_\mu}-m^2_{\nu_s}\approx m^2_{\nu_\mu}-m^2_{\nu_e}
\ga 4$ eV$^2$, the $\nu_\mu,\,\nu_\tau\rightleftharpoons\nu_s$ mixing
should also be extremely small.

Therefore, the mixing structures in these different cases of the
hierarchical four-neutrino schemes can be very disparate. Potentially,
unless the inter-doublet active-active and active-sterile mixings are
very asymmetric, BBN considerations demand a vacuum \lq\lq just
so\rq\rq\ solution to the solar neutrino problem and an LSND solution
with $m^2_\mu-m^2_e\approx 1$ eV$^2$ and $\sin^22\theta_{\mu e}
\approx 10^{-3}$ in the two-doublet hierarchical mass scheme.
Alternatively, future $\mue$ experiments such as BooNE could be able
to place significant constraints on functions of $\theta_{12},
\theta_{13}, \theta_{02}$ and $\theta_{03}$.  This is a tantalizing
result.  Realizing it depends on the veracity of the particular BBN
calculations we have adopted, but it is clear that the stakes are high.

X.S., G.M.F. and K.A. acknowledge partial support from
NSF grant PHY98-00980 at UCSD. K.A. wishes to acknowledge a NASA GSRP
Fellowship.

\newpage
\begin{table}
\caption{BBN Limits on the inter-doublet active-sterile
mixing in the two-doublet neutrino scheme.\label{table1}}
\vskip 0.5cm
\begin{tabular}{ccc}
$m^2_{\nu_\mu}-m^2_{\nu_e}$ & Type of Solar Neutrino Solution & Limit
on Effective $\sin^22\theta_{\mu s}$ and $\sin^22\theta_{\tau s}$
\\\tableline \multirow{2}{14mm}{$\ga 4$ eV$^2$} & SMA MSW &
\multirow{2}{14mm}{$\la 10^{-10}$} \\ & Vacuum & \\\hline
\multirow{2}{14mm}{$\la 4$ eV$^2$} & MSW & {$\la 10^{-10}$} \\ &
Vacuum & $\la 10^{-4}{\rm\, eV}^2\,\left(
m^2_{\nu_\mu}-m^2_{\nu_e}\right)^{-1}$ \\
\end{tabular}
\end{table}

\newpage

\newpage
\begin{figure}
\centertexdraw{
\drawdim truecm \linewd 0.04
\arrowheadtype t:F

\move(1 3)   \rlvec(2 0)
\move(1 3.07) \rlvec(2 0)
\move(1 3.14) \rlvec(2 0)

\move (2 4.9)
\ravec(0 0.97)

\textref h:C v:C
\htext(2 4.5){$\delta m^2_{\rm LSND}$}

\move (2 4.1)
\ravec(0 -0.9)

\lpatt(0.2 0.17)
\move(1 6)   \rlvec(2 0)

\textref h:C v:B
\htext(2 6.7){\small Scheme I}

\textref h:R v:C
\htext(.8 3.03){$\nu_e, \nu_\mu, \nu_\tau$}
\htext(.8 6){$\nu_s$}

\lpatt()
\move(6 3)   \rlvec(2 0)
\move(6 3.1) \rlvec(2 0)

\move (7 4.9)
\ravec(0 0.97)

\textref h:C v:C
\htext(7 4.5){$\delta m^2_{\rm LSND}$}

\move (7 4.1)
\ravec(0 -0.9)

\lpatt(0.2 0.15 0.01 0.15)
\move(6 6)   \rlvec(2 0)
\move(6 6.1)   \rlvec(2 0)

\textref h:C v:B
\htext(7 6.7){\small Scheme II}

\textref h:R v:C
\htext(5.8 3.03){$\nu_e, \nu_\tau$}
\htext(5.8 6){$\nu_\mu, \nu_s$}

\move(11 3)   \rlvec(2 0)
\move(11 3.1) \rlvec(2 0)

\lpatt()
\move (12 4.9)
\ravec(0 0.97)

\textref h:C v:C
\htext(12 4.5){$\delta m^2_{\rm LSND}$}

\move (12 4.1)
\ravec(0 -0.9)

\move(11 6)   \rlvec(2 0)
\move(11 6.1)   \rlvec(2 0)

\textref h:C v:B
\htext(12 6.7){\small Scheme III}

\textref h:R v:C
\htext(10.8 3.03){$\nu_e, \nu_s$}
\htext(10.8 6){$\nu_\mu, \nu_\tau$}

}
\vskip 2.5 cm
\caption[]{\small The three general mass heirarchies discussed.  In
all cases, the mass splittings correspond to $\delta m^2_{\rm
LSND}\sim 1\,{\rm eV^2}, \delta m^2_{\rm atm} \sim 10^{-3}\,{\rm
eV^2}$, and $\delta m^2_{\rm solar}\sim 10^{-5} (10^{-10})\,{\rm
eV^2}$ for the MSW (vacuum) solar solutions.  Scheme I, its mirror
($m_{\nu_e},m_{\nu_\mu},m_{\nu_\tau} > m_s$) and Scheme II (and its
mirror) have previously been ruled out by BBN.  In this paper, we
consider constraints on mass heirarchies and mixings in Scheme III.
\label{schemes}
}
\end{figure}

\newpage

\newpage

\end{document}